\documentclass[onecollarge]{article}       

\usepackage{graphicx}
\usepackage[ansinew]{inputenc}
\usepackage{hyperref}

\title{The experimental failure of macroscopic determinism: the case of an electrocardiogram}

\author{ R. Lapiedra\footnote{Corresponding author: R. Lapiedra. Departament d'Astronomia i Astrofísica, Universitat de València, 46100 Burjassot, Spain. Tel.: +34-963543077. Fax: +34-963543084.  e-mail: ramon.lapiedra@uv.es}\ , F. Montes \\
Universitat de València. Spain}

\date{}

\begin{document}

\maketitle

\begin{abstract}
Even if never elucidated, the question of determinism is a standing question along the history of human thinking. A physical system evolves in a deterministic way if its future is completely determined once we have fixed some present characteristics of it, i.e., its initial conditions. The problem addressed in the present paper is to test determinism in the macroscopic domain. By imposing a very plausible ``separability'' assumption, we prove that determinism enters in contradiction with the recorded outcomes of a given electrocardiogram. The interest of this result comes from the fact such a basic idea as determinism has never been experimentally tested up to now in the macroscopic domain, and as far as we know not even in the quantum domain.

\emph{Keywords}: Macroscopic determinism, Time Bell-like inequalities, Experimental failure of determinism
\end{abstract}

\section{Introduction}
\label{intro}
It is well known that realism, the previous existence of some hidden variable values determining any given outcome, has been experimentally tested in the quantum domain \cite{aspect82}, \cite{aspect99}, \cite{rowe01}. The experimental result is that, according to quantum mechanics predictions \cite{bell65}, \cite{clauser69}, what is called local realism has to be discarded, although some loopholes remain still open \cite{santos05}. Here, ``local'' means that the corresponding reality -the hypothetical so-called hidden variables- cannot involve correlations among its distant enough parts, i.e., among the parts unable to interchange signals because of the relativistic limit of interaction propagation, i.e., because of the limit of the speed of light.

Nevertheless, what we will test in the present paper is not realism, but determinism. In order to catch the difference between both concepts, consider two consecutive self-outcomes of a given isolated system. Imagine that we have some reality behind any outcome (realism). Let us consider the reality which is behind the first outcome. Then, it could happen that the same outcome erases this reality such that the reality which is supposed to be behind the second outcome is a different one (this is just what happens with the quantum state collapse in the measurement process). In such a case, we would have realism, but not necessarily determinism: in order to have determinism some initial conditions must exist, the same for all the system evolution. This is what is assumed to happen, implicitly at least, in classical physics.

But, letting aside the interest of determinism as such, what could be the remaining interest of testing determinism when realism has already been tested with the resulting rejection of local realism?  There are two reasons for this interest: in the first place, in the present paper we test macroscopic determinism by dealing with a very ordinary macroscopic system, and such a macroscopic test has never been performed as long as we know; in the second place, in the present paper framework, the locality condition is a too hard constraint since, even accepting the speed of light limit, as it should be, we could have some of the above precluded non local correlations by simply allowing the system to evolve in an appropriate deterministic way (a possibility to which Bell \cite{bell81} referred as a ``mind boggling … conspiratorially entangled'' world). Thus, what is really at stake behind the question of realism is just determinism and this is why we directly consider it here.

\section{The Case of an Electrocardiogram} \label{electro}
More precisely, we have considered the record of the successive times corresponding to the so called R self-responses (the typical sharp peaks) of an electrocardiogram. The file, together with the code used in the analysis that follows it (see the Appendix) contains 4340 recording times, from an anonymous patient's night hour record, whose mean time interval between two successive self-responses is $t_M = 829$ milliseconds. Then, we consider the time succession $\{t_n=3n t_M\}$, with $n=0,1,2,3,\ldots, 1446$. Let us consider too three time intervals $\tau_a, a=1,2,3$, such that $\tau_1\leq \tau_2\leq \tau_3$, with $\tau_3\leq 3t_M$. From the file of the above recording times, we will define a dichotomic (i.e., two-valued) function $U(t_n+ \tau_a)=\pm 1$ in the following way: If $\tau_a\leq  t_M$, $U$ takes the value +1 if during the time interval $(t_n, t_n+\tau_a)$ at least a response R appears, and the value -1 if it does not appear during this interval. Similarly in the two remaining cases:  $t_M\leq \tau_a \leq 2t_M$  with the interval $(t_n+t_M,tn+\tau_a)$, and $2t_M\leq \tau_a\leq 3t_M$ with the interval $(t_n+2t_M,t_n+\tau_a)$.

Then, imagine that for each time $t_n$ we choose at random two of the three obtained values $U(t_n+\tau_a)$. Let us consider the three corresponding correlation functions
\begin{equation}\label{eq1}
<U(a)U(b)>\equiv \displaystyle{\frac{1}{N_{ab}}\sum U(t_{n(ab)}+\tau_a)U(t_{n(ab)}+\tau_b)}, \ \ \   a<b; \ \   a,b =1,2,3,
\end{equation}
where $\{t_{n(ab)}\}$ is the sub ensemble of the time succession $\{t_n\}$ corresponding to the random selected pair $(\tau_a, \tau_b)$. The summation is over the $N_{ab}$ times $t_{n(ab)}$ and the addition of the three very similar numbers  $N_{ab}$ will give $N$.

In order to test it, let us assume determinism for the appearance of the successive R responses and so determinism for the physical system which is behind them. This determinism will be assumed, not for the heart producing the electrocardiogram, which is obviously a non isolated system, but for some ``enlarged system'' (the heart plus its actual environment: the entire human body, and even the patient's room and beyond) which can be considered isolated o nearly isolated. This means, in particular, that some initial conditions, $\Lambda_0$, referred to some initial time, $t_0$, exist, such that the above function $U$ can be written more explicitly $U(t_{n(ab)}+\tau_a,\Lambda_0)$. The initial conditions $\Lambda_0$ will collect the values of some variables of the ``enlarged system'' at this initial time. Notice that we also could write $U(\tau_a,\Lambda_{n(ab)})$ with $\Lambda_{n(ab)}$ denoting the new initial conditions corresponding to the times $t_{n(ab)}$ when these times are taken as new initial times. Obviously, given the original initial condition $\Lambda_0$ and the new initial time $t_{n(ab)}$, there is a unique initial condition $\Lambda_{n(ab)}$, but this uniqueness is not necessarily valid the other way around: the same value of $\Lambda$ could correspond to different initial times $t_{n(ab)}$ with the same $\Lambda_0$. Then, the correlation functions of (\ref{eq1}) can be written as
\begin{equation}\label{eq2}
<U(a)U(b)>\equiv \displaystyle{\sum p_{ab}(\Lambda_{n(ab)}) U(\tau_a,\Lambda_{n(ab)})U(\tau_b,\Lambda_{n(ab)})}
\end{equation}
where the summation is now over all different $\Lambda_{n(ab)}$  and we have introduced  the normalized probabilities $p_{ab}(\Lambda_{n(ab)})$ since the different values of $\Lambda_{n(ab)}$ will not appear, in principle, with the same frequency.

\section{Proving some Time Bell-like Inequalities} \label{bell}
Then, we make the following ``statistical separability assumption'': the history of  ``enlarged system'' (for example, the times in which the R responses appear), is statistically independent of the times $t_{n(ab)}$ selected by the authors. This means that in (\ref{eq2}) we can drop the index $ab$ affecting the probabilities $p_{ab}$, and the index $n(ab)$ affecting the $\Lambda$ values $\{\Lambda_{n(ab)}\}$. Thus, we can write (\ref{eq2}) as
\begin{equation}\label{eq3}
<U(a)U(b)>\equiv \displaystyle{\sum p(\Lambda) U(\tau_a,\Lambda)U(\tau_b,\Lambda)}
\end{equation}

Dropping the last index, $n(ab)$, needs some further comment. Notice first that we could write (\ref{eq3}) if the three sets $\{\Lambda_{n(ab)}\}$, irrespective of the order its elements,  were the same set, let us say $\{\Lambda\}$. Trying to ensure this, we could assume that the history of our ``enlarged system'' is completely independent of the ulterior choose of the times $t_{n(ab)}$ (of course, though stronger, this plausible assumption does not enter in contradiction with the above statistical independence). The problems is that, even with this full independence, it could happen that the three sets $\{\Lambda_{n(ab)}\}$ were not the same set: i.e., given an element of one of these three sets, it could happen that this element did not belong to some one of the other two sets. This could happen if the elements of $\{\Lambda_{n(ab)}\}$ did never repeat themselves. Nevertheless, the function $U(\tau_a,\Lambda_{n(ab)})$ is a two valued function. This means that its values repeat largely themselves. Thus, from the full independence we can expect that given any function value, $U(\tau_a,\Lambda_{n(ab)})$, there always will exist a value $\{\Lambda_{n(ac)}\}$ such that $U(\tau_a,\Lambda_{n(ac)})=U(\tau_a,\Lambda_{n(ab)})$, that is, such that we can write both as $U(\tau_a,\Lambda)$ with $\Lambda$ belonging to the above common set $\{\Lambda\}$, all which leads to (\ref{eq3}). However, in order to write (\ref{eq3}) we do not need to make such a natural assumption: it is enough to merely assume directly, as we have done, the less restrictive condition (\ref{eq3}) itself because of the assumed statistical independence.

From the two assumptions, determinism and statistical separability, we next prove the following ``time Bell-like inequalities'':
\begin{equation}\label{eq4}
D \equiv |<U(1)U(2)>- <U(1)U(3)>| +<U(2)U(3)> \leq 1,
\end{equation}
where the vertical bars mean taking absolute values.
In order to prove inequalities (\ref{eq4}) we will mimic here the proof of the original Bell's inequalities \cite{bell65}. Having in mind (\ref{eq3}), i.e., having in mind the assumed statistical independence, let us consider the difference
\begin{equation}\label{eq5}
<U(1)U(2)>-<U(1)U(3)>=\displaystyle{\sum p(\Lambda) U(\tau_1,\Lambda)U(\tau_2,\Lambda)}-\displaystyle{\sum p(\Lambda) U(\tau_1,\Lambda)U(\tau_3,\Lambda)}.
\end{equation}

Notice that a similar statistical independence is implicitly assumed when proving the original Bell's inequalities \cite{bell65} and further similar inequalities: there, the hidden variables, $\lambda$, are assumed to have a random distribution independent of the successive chose of the different pairs of measurement directions. A similar statistical independence is also assumed in \cite{legget08}, where it is called  the ``induction postulate''.

Having in mind that $U(\tau_a,\Lambda)^2 =+1$, inequality (\ref{eq5}) becomes
\begin{equation}\label{eq6}
<U(1)U(2)>-<U(1)U(3)>=\displaystyle{\sum p(\Lambda) U(\tau_1,\Lambda)U(\tau_2,\Lambda)[1-U(\tau_2,\Lambda)U(\tau_3,\Lambda)]} ,
\end{equation}

Then, taking absolute values,
\begin{equation}\label{eq7}
|<U(1)U(2)>-<U(1)U(3)>| \leq \displaystyle{\sum p(\Lambda) [1-U(\tau_2,\Lambda)U(\tau_3,\Lambda)]} ,
\end{equation}
that, according to $\sum p(\Lambda)=1$, becomes inequality (\ref{eq4}), as we wanted to prove.

Similar inequalities have been proved in the literature on the subject under several assumptions different from determinism. One of these assumptions (the non-invasive measurability) \cite{legget08,legget85} being very dubious and the other ones (joint realism \cite{lapiedra06} and counterfactual macroscopic definiteness \cite{legget08})  unnecessarily strong. In \cite{legget07}, it is claimed that inequalities (\ref{eq4}) have been proved, in the microscopic domain, from the assumption of determinism plus non contextuality. Something similar is claimed in \cite{lapiedra09}, without explicitly requiring the separability assumption. But, as we have just shown, the right proof of inequality (\ref{eq4}) needs this last assumption, besides determinism.

\section{Testing the Violation of the Time Bell-like Inequalities. Conclusions} \label{conclus}
The main result of the present paper is the violation of inequality (\ref{eq4}) by the recorded data of a given electrocardiogram. This entails the failure of determinism as far as the evolution of the above ``enlarged system'', including the corresponding heart, is concerned, and so the failure of determinism in the natural world. Certainly we could elude this conclusion, but only at the high price of allowing the existence of some convenient correlations between the recorded data and our arbitrarily differed choice of the times $t_n$ and $\tau_a$.

The correlation functions $<U(a)U(b)>$ have been estimated using (\ref{eq1}) for different combinations of $\tau_a, \tau_b$, $a<b$, $(a,b)=1,2,3$. These are the $6^3=216$ combinations resulting when $\tau_1$ is chosen among the six times $(300, 400, \ldots , 800)$, and similarly for $\tau_2$ among $(t_M + 300, t_M + 400, \ldots t_M +800)$ and $\tau_3$ among $(2t_M + 300, 2t_M + 400,\ldots , 2t_M + 800)$. For 19 of these 216 combinations inequality (\ref{eq4}) is violated. Its greatest violation corresponds to the combination $\tau_1=800$, $\tau_2= t_M +600$ and $\tau_3= 2t_M +300$ for which the value of $D$ in inequality (\ref{eq4}) becomes $D = 1,12$.

\begin{figure}[h]
\begin{center}
  \includegraphics[width=0.5\textwidth]{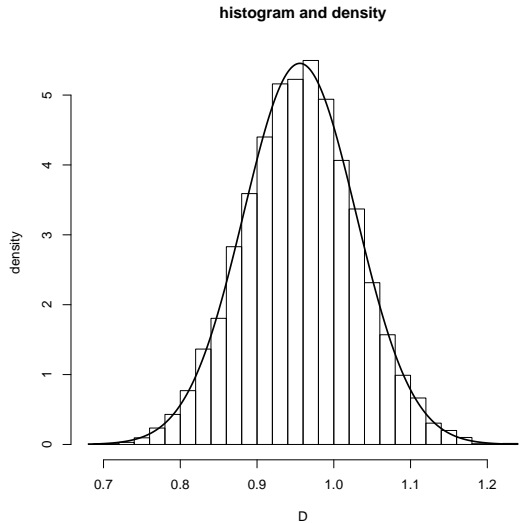}

\caption{Histogram and adjusted Normal density for the $10^4$ repetitions of $\tau$'s combination giving maximum $
D$.} \label{fig1}
\end{center}
\end{figure}

It is necessary to remember that this violation value is produced by a given random choice of the sequence of pair times $(\tau_a, \tau_b)$, $a<b$, $(a,b)=1,2,3$, for each element of $\{t_n=3n t_M \}$, with $n=0,1,2,3, \ldots, 1446$. There are an enormous number, $3^{1446}$, of such possible choices. This means that the probability of getting the above result, jointly with the other 18 violations, by mere chance is very small, as we discuss below. To this purpose, we have generated $10^4$ different random sequences for the same combination of $\tau$'s: $\tau_1=800$, $\tau_2= t_M +600$ and $\tau_3= 2t_M +300$. A total of 2251 violations, ranging from 1,01 to 1,24, were encountered confirming that the original result is not due to chance in the particular selection of one of the above $3^{1446}$ different sequences. A Kolmogorov-Smirnov test has been performed for testing the normality of these $10^4$ observations. The statistic of the test is 0,0042 with a p-value of 0,4995. So the normality of the data can be accepted (Fig. \ref{fig1}).

We now confirm, with a deeper analysis, that the above violation is not due to a particular choice of the above combination of $\tau_a$, by significantly shifting this combination $\tau_1=800$, $\tau_2= t_M +600$ and $\tau_3= 2t_M +300$, in its neighbourhood. Each time, we have replaced some of the $\tau_a$ by $\tau_a \pm 20$. The resulting new combinations have been explored as described above, i.e., by randomly generating each time $10^4$ random sequences. This complementary study confirms the original violation of $D\leq 1$. In particular, for the combination $\tau_1=780$, $\tau_2= t_M +620$ and $\tau_3= 2t_M +320$, we have even found a greater violation than before, corresponding to $D = 1,28$.

\section{Final Considerations} \label{final}
As we have just explained, this violation entails that the R responses of our electrocardiogram have not been generated in a deterministic way, unless we are disposed to leave an assumption as unavoidable as the statistical separability. Notice that this assumption does not invalidate determinism: our choice of the times $t_n$ and $\tau_a$ could be perfectly deterministic, if we want. All is needed here is that this arbitrarily delayed choice be independent, or which is less restrictive, statistically independent, of the previously recorded R responses.

Notice that we have not needed to assume locality: our assumed initial conditions are not required to be local. Assuming locality would have been no consistent in the present case, since as we have pointed above one could have non local realism by allowing the system to evolve in a deterministic way, and it is just determinism what we have tentatively assumed here in order to test it. At the same time, the inefficient detection loophole \cite{rowe01} is not present here, since all the self-responses, R, of our electrocardiogram have been considered in our analysis.

Let us make another remark: the expression for $<U(a)U(b)>$ given in (\ref{eq3}) says that we deal with a common probability space, i.e., we use the same probability function, $p(\Lambda)$, and we can put $\Lambda$ instead of $\{\Lambda_{n(ab)}\}$, all along the history of the system. This common space exists because of the particular statistical independence entailed by the separability assumption, which has allowed us to drop the index $ab$ and $n(ab)$ from the probabilities $p_{ab}$ and $\{\Lambda_{n(ab)}\}$, respectively, in (\ref{eq2}). However, could not it be the case that successive new initial conditions were incorporated to the actual initial conditions of the R responses along the history of the heart and its assumed environment (the ``enlarged system''), these new initial conditions not being before present because of the relativistic limit of propagation of interactions? If this were the case, it would not be evident that such a common probability space exists. But, in the present case, there is not such a problem since, because of determinism, there is no exterior effects on this enlarged system, containing the heart and its self-responses (a system enlarged as to cover the entire universe if needed), neither there is really anything new happening in this system since everything what happens there comes from the initial conditions. As a result we must have a common probability space for the $\Lambda$ values in $U(\tau_a,\Lambda)$ and $p(\Lambda)$, such that we be able to write the correlation functions $<U(a)U(b)>$ in the way we have done in (\ref{eq3}). The question of the possible lack of a common probability space has also been raised \cite{hess05} in the similar scenario of the seminal Bell's inequalities \cite{bell65}, where, as we have remarked above, assuming the existence of this space is actually an implicit, even if plausible, hypothesis that we do not need to make here because of the determinism assumption.

Finally, we do not have at present any theory leading to the failure of determinism, out of quantum mechanics. Thus, we could speculate that the reported violation of macroscopic determinism could be the macroscopic magnification of the quantum indeterminism of the microscopic constituents of the heart and its environment. Furthermore, we could ask ourselves if this hypothetical magnification comes from the fact that we were dealing with living stuff, or more generally it would be expected to happen every time that we have some sort of dichotomist self-response in an, living or not, isolated system. All these questions would deserve some future work.

\section*{Acknowledgements}
We thank Armando Pérez for useful discussions and Lluís Garcia Sevilla for suggesting us to try with an electrocardiogram. We also thank Drs. Àngel Llàcer and Vicente Ruiz for providing us with the electrocardiogram file. R.L. was supported by the Spanish Ministerio de Ciencia e Innovación MICIN-FEDER project No. FIS2009-07705. F.M. was supported by the Spanish Ministerio de Ciencia e Innovación project MTM2008-05152.

\section*{Appendix}
Readers can recover the data file and can reproduce the analysis results, described in the article, downloading the files \textsf{ecgn.txt} and file \textsf{analysisR.txt} from http://www.uv.es/montes/electrocardiogram . The first file contains the data corresponding to the electrocardiogram with individual and accumulative response times, the second one is an \emph{R} \cite{Rcode} code allowing to reproduce the analysis performed. This code provides three output files:
\begin{description}
  \item[\textbf{output1.tx}:] the results generated by the 216 combinations of  $\tau_1$,  $\tau_2$ and  $\tau_3$. The columns are, in this order,  $\tau_1$,  $\tau_2-t_M$,  $\tau_3-2t_M$, $D$, $n_{12}$, $n_{13}$ and $n_{23}$
  \item[\textbf{output2.txt}:] the results generated by $10^4$ random repetitions obtained for the combination of  $\tau$'s providing the maximum $D$ among the 216 of the above file. The file has the same structure as \textbf{output1.txt} and, obviously, $\tau_1$,  $\tau_2$ and  $\tau_3$ have the same value along the $10^4$ repetitions.
  \item[\textbf{summary.txt}:] a self explicative file summarizing the analysis, including also the result of the Kolmogorov-Smirnov test for testing the normality of the $10^4$ $D$ values in the above file.
\end{description}

It is possible to obtain an exact copy of our results if the same random seed is used at the beginning of the \emph{R} code. In our case this value was 12345.

\end{document}